\documentclass[%
 reprint,
 amsmath,amssymb,
 aps,
]{revtex4-1}

\usepackage{multirow}
\usepackage{listings}
\usepackage{graphicx}
\usepackage{textcomp}
\usepackage{dcolumn}
\usepackage{bm}
\usepackage{hyperref}
\usepackage{xcolor}
\hypersetup{
    colorlinks,
    linkcolor={red!50!black},
    citecolor={green!50!black},
    urlcolor={blue!50!black}
}

\renewcommand{\>}{\rangle}
\newcommand{\<}{\langle}

\definecolor{codegreen}{rgb}{0,0.6,0}
\definecolor{codegray}{rgb}{0.5,0.5,0.5}
\definecolor{codepurple}{rgb}{0.58,0,0.82}
\definecolor{backcolour}{rgb}{0.95,0.95,0.92}
 
\lstdefinestyle{mystyle}{
    backgroundcolor=\color{lightgray!30},   
    commentstyle=\color{codegreen},
    keywordstyle=\color{magenta},
    numberstyle=\tiny\color{codegray},
    stringstyle=\color{codepurple},
    basicstyle=\footnotesize,
    breakatwhitespace=false,         
    breaklines=true,                 
    captionpos=b,                    
    keepspaces=true,                 
    numbers=left,                    
    numbersep=5pt,                  
    showspaces=false,                
    showstringspaces=false,
    showtabs=false,                  
    tabsize=2,
}
 
\begin{document}


\title{Distributed Memory Techniques for Classical Simulation of Quantum Circuits}

\author{Ryan LaRose}
 \email{laroser1@msu.edu}
\affiliation{Department of Computational Mathematics, Science, and Engineering, Michigan State University.}
 \altaffiliation[Also at ]{Department of Physics and Astronomy, Michigan State University.}

\date{\today}

\begin{abstract}
In this paper we describe, implement, and test the performance of distributed memory simulations of quantum circuits on the MSU Laconia Top500 supercomputer. Using OpenMP and MPI hybrid parallelization, we first use a distributed matrix-vector multiplication with one-dimensional partitioning and discuss the shortcomings of this method due to the exponential memory requirements in simulating quantum computers. We then describe a more efficient method that stores only the $2^n$ amplitudes of the $n$ qubit state vector $|\psi\>$ and optimize its single node performance. In our multi-node implementation, we use a single amplitude communication protocol that maximizes the number of qubits able to be simulated and minimizes the ratio of qubits that require communication to those that do not, and we present an algorithm for efficiently determining communication pairs among processors. We simulate up to 30 qubits on a single node and 33 qubits with the state vector partitioned across 64 nodes. Lastly, we discuss the advantages and disadvantages of our communication scheme, propose potential improvements, and describe other optimizations such as storing the state vector non-sequentially in memory to map communication requirements to idle qubits in the circuit. 
\end{abstract}

\pacs{Valid PACS appear here}
\maketitle



\section{\label{sec:intro}Introduction}

Quantum computers excel at certain computational tasks where classical computers struggle. Shor's algorithm for the quantum Fourier transform (QFT) \cite{shor-qft}, which performs exponentially faster than the classical fast Fourier transform, was the first to ignite the field and spark interest in what else quantum computers can do efficiently. The QFT is now used as a subroutine in several other quantum algorithms including factoring, estimating eigenvalues, and computing the discrete logarithm \cite{nielsen-and-chuang}. Quantum computers have been shown to have promising applications in quantum chemistry \cite{feynmann-original,qsim-nsf-report,qsim-georgescu,plane-wave-basis} for simulating the Schr\"{o}dinger equation
\begin{equation*} \label{eqn:schrodinger}
i \hbar \frac{\partial \Psi}{\partial t} = H \Psi 
\end{equation*}
of electronic Hamiltonians
{\small
\begin{equation*} \label{eqn:electronic-hamiltonian}
H = - \sum_{i} \frac{\nabla_i^2}{2} - \sum_{i, j} \frac{z_j}{|R_j - r_i|} 
    + \sum_{i < j} \frac{1}{|r_i - r_j|} + \sum_{i < j} \frac{z_i z_j}{|R_i - R_j|}
\end{equation*}
}
where $r_i$ is the position of the $i$th electron and $(R_i, z_i)$ are the position and charge of the $i$th nucleus; in optimization \cite{quant-anneal-vadim-dykman} for exploiting quantum tunneling for quickly finding the ground state energy (solution) of combinatorial optimization problems; and machine learning \cite{lloyd-qml-nature,alan-qml,lloyd-qml-quantum-gradient-descent,lloyd-qml-svd} for, under certain conditions, performing tasks such as principal component analysis and singular value decomposition 
\begin{equation*} \label{eqn:svd}
	A = U \Sigma V^*
\end{equation*}
exponentially faster than classical machines. 

Although quantum computers have been largely theoretical for the last two decades, efforts in industry and academia alike are materializing in real world quantum computers. The ``IBM Quantum Experience'' offers a 5 qubit and 16 qubit quantum computer available to use through a cloud-based queue \cite{ibmqx-website,ibmqx-5qubits,ibmqx-16qubits} and D-Wave has for sale a 2048 qubit machine based on the adiabatic model of quantum computation \cite{dwave-website}. These machines are either too small to perform significant computations or too large to be truthfully quantum \cite{d-wave-rebuttle}, but they represent big strides in moving towards larger-scale quantum computers that achieve ``quantum supremacy'' by performing a task efficiently that no classical computer can \cite{google-qsuprem}. The date of arrival of such a quantum machine is predicted by several to be within the next five to ten years \cite{google-qsuprem,qfive-years}.

Until that time, for the purposes of testing algorithms and performing research in quantum information science, it is necessary to have a means of simulating a quantum computer. Since quantum computers operate on linear algebra, which classical computers can do, it is possible to simulate them on classical computers. However, it is not necessarily possible to simulate them efficiently due to exponential growth in memory requirements as the number of qubits increases. It thus becomes an interesting problem in classical computer science to develop techniques to handle this exponential ``explosion'' \cite{feynmann-original}; it is this problem we focus on in this paper. 

Several quantum computing software packages have recently been developed that use distributed memory techniques on the world's best supercomputers to simulate numbers of qubits nearing the quantum supremacy threshold \cite{julich-long-thesis,qhipster,45qubit-simulation,QX}. Other software uses information compression schemes and efficient data structures to achieve similar results \cite{adv-sim-compression-overview-2017,qtorch}. Our contributions in this paper are the following: we implement a single amplitude communication scheme in the commonly used ``state vector partitioning'' method, in which the $n$ qubit quantum state $|\psi\>$ of $2^n$ complex amplitudes is partitioned among $2^k$ processors, and test its performance. We use the amplitude updating procedure outlined in \cite{julich-long-thesis} and additionally present an efficient algorithm for determining the processors that need to communicate with one another in the multi-node implementation. We discuss how our communication scheme, compared to others, allows more qubits to be simulated and maximizes the ratio of qubits that do not require communication to qubits that do, thereby decreasing overall simulation time of a circuit at the cost of high single gate application time for the qubits that require communication in updating. Lastly, we discuss how storing the state vector non-sequentially can map qubits requiring communication to ``idle'' qubits in the circuit as a means of mitigating the high gate application time of our communication protocol for qubits that need communication in updating.

The rest of the paper is organized as follows. For self-containment we include a brief discussion of the fundamentals of quantum computation. After mentioning the hardware and software used in our implementation and testing, we demonstrate the exponential memory problem by applying a distributed memory parallel matrix vector multiplication, allowing us to simulate up to 17 qubits. We then describe in detail the state vector partitioning method and present our communication protocol. This method enables us to simulate up to 30 qubits on a single node and 33 by partitioning the state vector amongst several processors. We report single gate timing statistics for each method and discuss the advantages and disadvantages of our implementation. We end by proposing improvements to our method and proposing non-sequential storage of the state vector.


\section{\label{sec:quantum-computing}Quantum Computing Basics}

\subsection{\label{subsec:basic-principles}Quantum Information Processing}

Classical computers store information in binary digits (\textit{bits}) which can take the value of 0 or 1. This information is manipulated by logic gates such as \texttt{AND}, \texttt{XOR}, and \texttt{NOT} which take in 2 bits as input and produce one bit as output according to well-defined rules. The basis for classical computation is contained in these two principles.

Quantum computers operate on different ideas inspired by the 20th century advent of quantum physics. Information is no longer stored in a binary variable but rather a two level quantum system\footnote{It is also possible to store information in a $d$-level system with $d > 2$. These bits of information are known as \textit{qudits} but will not be discussed in this paper.}. In reference to classical computing, these are known as \textit{quantum bits}, or \textit{qubits}, and can take the values $| 0 \>$ and $|1\>$. Qubits are manipulated by unitary operators such as the Pauli matrices $X$, $Y$, $Z$ and the Hadamard gate $H$, to be discussed more in depth shortly. A knowledge of quantum mechanics is helpful in quantum computing, but all that is required is a foundation in linear algebra.

\begin{figure}
	\centering
    \includegraphics[scale=0.7]{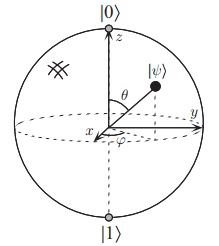}
    \caption{A graphical representation of the Bloch sphere and an arbitrary qubit $|\psi\>$ \cite{nielsen-and-chuang}. Angles $\theta$ and $\phi$ measured from the $z$ and $x$ axes, respectively, are also equivalent ways to express qubits.}
    \label{fig:bloch-sphere}
\end{figure}

A quantum bit $| \psi \>$ is simply a column vector with complex coefficients. (The notation $| \cdot \>$ is due to Dirac and is called a ``ket''|anything enclosed in a ket is a column vector and the symbol that is enclosed is a label. The symbol $\<\cdot|$, called a ``bra,'' denotes the conjugate transpose.) The vectors
\begin{equation} \label{eqn:computational-basis-vectors}
    |0\> \equiv \left[ \begin{matrix}
    	1 \\
        0 \\
    \end{matrix} \right], \qquad 
    |1\> \equiv \left[ \begin{matrix}
    	0 \\
        1 \\
    \end{matrix} \right]
\end{equation}
form a basis for $\mathbb{C}^2$, known as the \textit{computational basis}, and as such any qubit can be expressed as a linear combination
\begin{equation} \label{eqn:arbitrary-qubit}
	| \psi \> = \alpha | 0 \> + \beta | 1 \> 
\end{equation}
for any $\alpha, \beta \in \mathbb{C}$ such that $|\alpha|^2 + |\beta|^2 = 1$. This stipulation is because wave-functions $\psi$ in quantum mechanics must be normalized: $|\psi|^2 = \psi^{*}\psi$ is interpreted as a probability distribution. Quantum bits thus are two-dimensional complex vectors of unit norm, meaning they sit on the unit sphere in $\mathbb{C}^2$ known as the \textit{Bloch sphere}. See Figure \ref{fig:bloch-sphere}.

A qubit is manipulated on a quantum computer by a unitary operator|a matrix $U \in \mathbb{C}^{2 \times 2}$ such that $U^{*} = U^{-1}$, where the star denotes the conjugate transpose $U_{ij}^{*} = \bar{U}_{ji}$. Just as bits and logic gates form the basis for classical computation, qubits and unitary operators (also called gates) form the basis for quantum computation. One gate, already mentioned, is the Pauli $X$ gate
\begin{equation} \label{eqn:pauli-x-gate}
X \equiv \left[ \begin{matrix}
	0 & 1 \\
    1 & 0 \\
\end{matrix} \right]
\end{equation}
which has the readily verified property that $X |0\> = |1\>$ and $X |1\> = |0\>$, thus making it the quantum analog of the classical \texttt{NOT} gate. Another gate to mention is the Hadamard gate $H$, for it leads to a phenomena not possible in classical computing. Observe that
\begin{equation} \label{eqn:hadamard-gate-definition}
H \equiv \frac{1}{\sqrt{2}} \left[ \begin{matrix}
	1 & 1 \\
    1 & -1 \\
\end{matrix} \right] 
\end{equation}
has the property that
\begin{equation} \label{eqn:hadamard-gate-action}
	H |0\> = \frac{|0\> + |1\>}{\sqrt{2}} \qquad \textrm{and} \qquad H |1\> = \frac{|0\> - |1\>}{\sqrt{2}},
\end{equation}
illustrating the principle of \textit{superposition}: If $|\psi_1\>$ is an allowable state and $| \psi_2 \>$ is an allowable state, then so is any linear combination $a |\psi_1\> + b |\psi_2\>$ so long as it's properly normalized. Thus, qubits can hold the value of both zero and one at the same time.

Superposition is one feature of quantum computing that allows information to be processed efficiently. The second is \textit{entanglement}, which governs how multiple qubits combine. For a system with $n$ qubits $| \psi_0 \>, ..., | \psi_{n - 1} \>$, the state of the whole system is determined by the $2^n$ dimensional tensor product
\begin{equation} \label{eqn:tensor-product-qubits}
    |\psi\> = | \psi_0 \> \otimes | \psi_1 \> \otimes \cdots \otimes | \psi_{n - 1} \>,
\end{equation}
commonly abbreviated as $|\psi\> = |\psi_0 \psi_1 \cdots \psi_{n - 1} \>$. For example, with two qubits $|\psi_0\> = \left[ \alpha_0 \ \beta_0 \right]^T$ and $|\psi_1\> = \left[ \alpha_1 \ \beta_1 \right]^T$, the state vector of the system is
\begin{equation} \label{eqn:tensor-product-2qubit-example}
	|\psi\> = | \psi_0 \> \otimes |\psi_1\> = 
    \left[ \begin{matrix}
    	\alpha_0 \\
        \beta_0
    \end{matrix} \right] \otimes \left[ \begin{matrix}
    	\alpha_1 \\
        \beta_1
    \end{matrix} \right]
    = 
    \left[ \begin{matrix}
    	\alpha_0 \alpha_1 \\
        \alpha_0 \beta_1 \\
        \beta_0  \alpha_1 \\
        \beta_0 \beta_1
    \end{matrix} \right]
\end{equation}
For each additional qubit, the dimension of the Hilbert space increases by a factor of 2. This exponential increase contains the power of quantum computation|and also the difficulty in classically simulating it. 

The single qubit gates discussed above combine in the same way as qubits to act on the state of the entire system. If the gate $U_i$ acts on qubit $| \psi_i \>$, $0 \le i < n$, then the unitary operator that acts on the whole system is the tensor product
\begin{equation} \label{eqn:tensor-product-gates}
U = U_0 \otimes U_1 \otimes \cdots \otimes U_{n - 1} \equiv \bigotimes_{i = 0}^{n - 1} U_i,
\end{equation}
and the state evolves according to 
\begin{equation} \label{eqn:circuit-level}
    | \psi ' \> = U |\psi\> = \left[\bigotimes_{i = 0}^{n - 1} U_i \right] \left[\bigotimes_{i = 0}^{n - 1} | \psi_i \> \right] = \bigotimes_{i = 0}^{n - 1} U_i |\psi_i\>.
\end{equation}

There are also multi-qubit gates in quantum computing that operate on more than one qubit such as the controlled not gate
\begin{equation*}
	\texttt{CNOT} \equiv \left[ \begin{matrix}
	1 & 0 & 0 & 0 \\
    0 & 1 & 0 & 0 \\
    0 & 0 & 0 & 1 \\
    0 & 0 & 1 & 0 \\
	\end{matrix} \right]
\end{equation*}
This operator acts on the four computational basis states and flips the second qubit (the \textit{target} qubit) only if the first qubit (the \textit{control} qubit) is one:
\begin{align*}
\texttt{CNOT} |00\> = |00\>, \ 
&\texttt{CNOT} |01\> = |01\>, \ \\
\texttt{CNOT} |10\> = |11\>, \ 
&\texttt{CNOT} |11\> = |10\>, \ 
\end{align*}
Other gates like the Toffoli gate, the controlled-controlled not gate, act on more than 2 qubits. For simplicity in our implementation and performance testing, we only consider single qubit gates.

\subsection{\label{subsec:quantum-circuits}Quantum Circuits}

\begin{figure}
	\centering
    \includegraphics[scale=0.8]{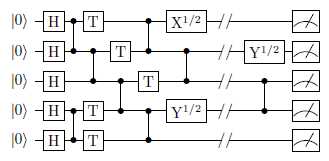}
    \caption{The layout of a quantum circuit. The qubit register is shown on the left and black lines show evolution in time. Gates are applied to the qubit as they appear. Multi-qubit gates, which in this circuit are all controlled $Z$ gates, affect more than one qubit and are drawn as vertical lines connecting the two qubits they operate on. The symbol on the far right of each qubit represent measurements. This circuit was used in \cite{google-qsuprem} to study the classical problem of efficiently simulating quantum circuits.}
    \label{fig:qcircuit-example}
\end{figure}

A quantum circuit consists of a sequence of gates applied to a group of qubits, commonly called a \textit{register}. The term \textit{circuit} is used only in analogy with classical computation; quantum information does not travel in loops of wires like classical bits do. A typical schematic diagram of a quantum circuit is shown in Figure \ref{fig:qcircuit-example} for five qubits. This particular circuit does no useful computation but was rather selected to make classical simulations as difficult as possible. 

Every quantum algorithm, such as the QFT, has a corresponding circuit. It is of great interest, particularly in simulating quantum computers, to know the circuit for an algorithm with the least number of gates. This is a theoretical optimization and one we do not consider in this paper, but an important one nonetheless.

Fundamental quantum circuits like the QFT, entanglement, and quantum teleportation are commonly used as benchmarks in evaluating the performance of quantum computer simulators \cite{adv-sim-compression-overview-2017}. In our testing, we focus solely on the time to apply a single gate to some qubit in the circuit, an equally important and frequently cited standard for evaluating performance. The step from gate application to circuit simulation is simply a matter of applying more gates.


\section{\label{sec:software-hardware-details}Hardware and Software Details}

We perform simulations on Michigan State University's Laconia system, a Top500 supercomputer whose clusters include 596 compute nodes connected by FDR/EDR Infiniband with more than 17,500 cores \cite{top500,msu-icer}. In particular, we use the intel16 cluster containing 320 total available nodes with 28 cores per node for a total of 8960 cores. On this cluster, 290 nodes have a maximum memory of 128 GB, 24 nodes have 256 GB, and 6 nodes have 512 GB. 

We use MPI (OpenMPI 1.6.5) for distributed memory computing and OpenMP 4.0 for parallelization among threads. Specific details on how parallelization is utilized is documented in the section of each implementation. For common quantum operations, we use Quantum++ (Q++) \cite{qpp}, an open-source C++11 quantum computing library. There have been a large amount of similar recent libraries implemented in several languages from Python \cite{qiskit,openqasm} to F\# \cite{liquid}. Q++ was used in this research for its simplicity, flexibility, and compatibility with parallel frameworks. 


\section{\label{sec:direct-approach}Direct Approach}


\subsection{\label{sec:direct-methodology}Methodology}

Subroutines in linear algebra like solving a system of equations are well-studied and used as benchmarks for supercomputing architectures \cite{linpack-benchmark}. Matrix vector multiplication is a common operation that is parallelized \cite{barlas-parallel-textbook}, and since applying a gate level in a quantum circuit boils down to matrix vector multiplication (\ref{eqn:circuit-level})
\begin{equation}
	| \psi ' \> = U | \psi \> ,
\end{equation}
this method is a natural first approach.


\subsection{\label{sec:direct-implementation}Implementation}

We use a one-dimensional matrix partitioning to carry out the matrix vector multiplication
{\footnotesize
\begin{lstlisting}[language=C]
for (i = 0; i < 2^n; i++) {
    temp = 0.0;
    #pragma omp parallel for \
    num_threads(NUM_THREADS) reduction(+:temp)
    for (j = 0; j < 2^n; j++) {
        temp += U[i * n + j] * psi[j];
    }
    psinew[i] = temp;
}
\end{lstlisting} }
in parallel for a circuit of $n$ qubits, storing the unitary $U$ as a long vector of $2^{n} \times 2^{n}$ elements. Our partitioning scheme of the matrix $U$ is shown in Figure \ref{fig:1dpartition}. Each processor must store a local copy of the state vector $|\psi\>$ which has $N = 2^n$ elements, so the lower bound on the total communication volume goes as $\Omega (pN)$ where $p = 2^{k}$ is the number of processors used and $\Omega$ has the usual meaning in computational complexity. Once each process has a partition of the unitary $U$ and a copy of the state vector $|\psi\>$, local computations are performed then the results are collected into the vector $|\psi '\>$. We utilize multiple threads in the inner \texttt{for} loop to speed up local computations on each node.

For a state vector of $n$ qubits with each element stored in single precision, $8 = 4 + 4$ bytes are required for the real and imaginary parts of the $2^n$ amplitudes. If we use $p = 2^k$ processors, each node stores $2^{n + 4}$ bytes for the state vector and an additional $2^{n - k + 4}$ bytes for the local partitioning of the matrix. With one node, the total memory requirement is thus $2 \cdot 2^{n + 4}$ bytes. For nodes with $128 \ \textrm{GB} = 2^{37}$ bytes on the MSU Laconia system, we can thus simulate approximately $n = 14$ qubits. In practice, since some memory is reserved for the OS and other tasks \cite{qhipster}, the maximum number of qubits able to be simulated cannot actually be reached. 

\begin{figure}
	\centering
    \includegraphics[scale=0.4]{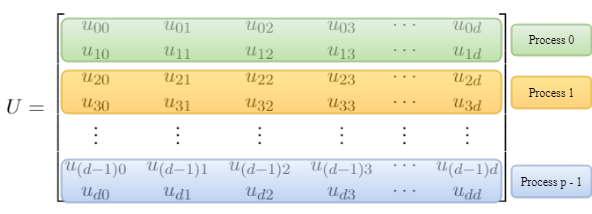}
    \caption{The one-dimensional partitioning method used for the parallel matrix vector multiplication approach. The matrix $U$ is stored as a linear array in memory|the two-dimensional representation here is for convenience in illustration. Here, the dimension of the matrix is $d = 2^n$ where $n$ is the number of qubits, and $p = 2^{k}$ processors are used where $k = 0, 1, 2, ...$ is an integer.}
    \label{fig:1dpartition}
\end{figure}

\begin{figure}
	\centering
    \includegraphics[scale=0.5]{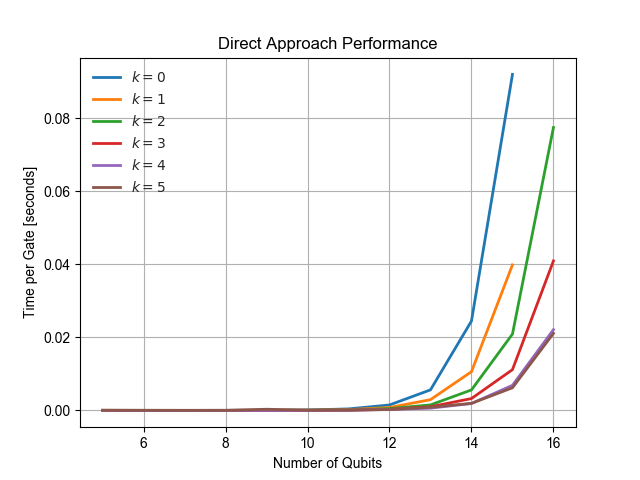}
    \caption{Performance of the parallel matrix-vector multiply implementation using $2^k$ processors (different color curves). The time per gate application is for the zeroth qubit in the circuit. As can be seen, most processors perform about the same in terms of time, but utilizing more processors enables more memory and thus a higher number of qubits that can be simulated.}
    \label{fig:1dpartition}
\end{figure}

A plot of the single gate application time of this method is shown in Figure \ref{fig:1dpartition}. Note that we restricted testing to random unitary matrices and we kept the entries in both the matrix and the unitary to be real-valued. Thus we are only required to store one double for each element and can simulate (in principle) four more qubits than the expected cap of 14. In practice, we are able to simulate up to 17 qubits by utilizing 32 processors. Since the unitary $U$ acts on all $n$ qubits, we divide the total elapsed time by the number of qubits in the circuit in reporting our results.


\subsection{\label{sec:2d-partition}2D Partitioning}

One may be tempted to implement a two dimensional partitioning of the unitary $U$ to decrease communication volume and overall computation time. Indeed a 2D partitioning would achieve these to some degree, but the limiting factor is the exponential memory requirement in the number of qubits. In the above implementation each process must store a local copy of the state vector $| \psi \>$, limiting the number of qubits to be simulated to $n = 14$ on nodes with 128 GB memory. With a 2D partitioning the memory requirements would decrease slightly since only a portion of $|\psi \>$ is stored on each processor, but each node would still have to store a portion of the $2^n \times 2^n$ unitary $U$. In the next section we implement a method that avoids having to store the entire matrix $U$.


\section{\label{sec:state-vector}State Vector Partitioning}


\subsection{\label{sec:state-vector-methodology}Methodology}

The approach of the several quantum circuit simulators \cite{julich-long-thesis,qhipster,45qubit-simulation} is to reduce memory requirements as much as possible by (i) not forming the $2^n \times 2^n$ unitary matrix $U$ and (ii) partitioning the state vector $| \psi \>$ among several processors. The product $|\psi ' \> = U | \psi \>$ is achieved by mimicking the behavior of the matrix vector multiplication, as follows.

Consider an arbitrary unitary matrix $Q_i \in \mathbb{C}^{2 \times 2}$ acting on the $i$th qubit in the circuit
\begin{equation}
	Q_i \equiv \left[ \begin{matrix}
		q_{11} & q_{12} \\
        q_{21} & q_{22} \\
	\end{matrix} \right] .
\end{equation}
As per (\ref{eqn:circuit-level}, the unitary acting on the entire state is given by
\begin{equation}
	U = \underbrace{I \otimes \cdots \otimes I }_{i - 1 \ \text{terms}}
        \otimes \ Q_i \otimes 
        \underbrace{ I \otimes \cdots \otimes I }_{n - i \ \text{terms}} .
\end{equation}
If qubits are uncoupled, one can simply apply the matrix $Q_i$ to the $i$th qubit to get the new state 
\begin{align}
	| \psi ' \> &= (I \otimes \cdots \otimes I \otimes Q_i \otimes I \otimes \cdots \otimes I) |\psi_0 \cdots \psi_{n-1}\>  \nonumber \\
    &= |\psi_0\> \otimes \cdots \otimes Q_i | \psi_i \> \otimes \cdots \otimes |\psi_{n-1}\>,
\end{align}
where $| \psi_i \> = [\alpha_i \ \beta_i]^T$ updates according to 
\begin{equation} \label{eqn:mtx-product-update-psi}
U_i | \psi_i \> = \left[ \begin{matrix}
    q_{11} \alpha_i + q_{12} \beta_i \\
    q_{21} \alpha_i + q_{22} \beta_i \\
\end{matrix} \right] .
\end{equation}
\begin{figure}
	\centering
    \includegraphics[scale=0.50]{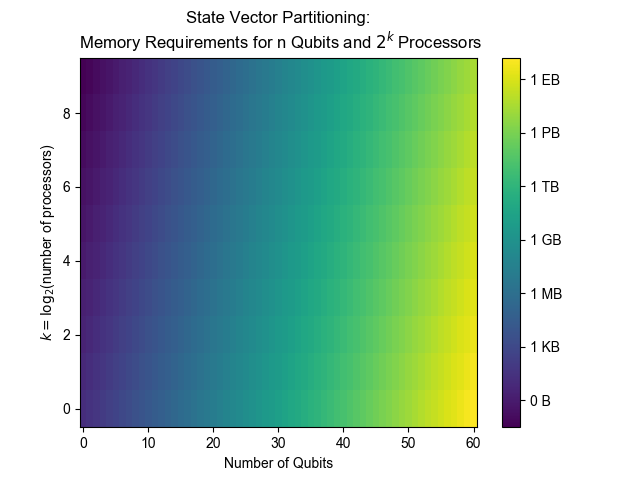}
    \caption{The single node memory requirements for $n$ qubits (horizontal axis) using $2^k$ processors (vertical axis) in the state vector partitioning method. Due to the OS and other tasks taking up memory, as well as leaving space on processors for communication (which depends on the communication protocol), some extra space is necessary|values in this figure can be regarded as lower bounds.}
    \label{fig:mem-requirements}
\end{figure}
It is a common operation in quantum computing to perform the Hadamard transform $H$ on each qubit in a circuit immediately after preparing the initial state $|0\>^{\otimes n}$ to get the Bell state $H |0\>^{\otimes n} = |+\>^{\otimes n} $: single qubit gates at the start of the circuit may take advantage of this optimization. However, it becomes tedious to implement for later levels in the circuit and it is easier, though less efficient, to form the entire state vector of $2^n$ amplitudes from the onset. 

Let the state vector be represented by amplitudes $\alpha$ indexed with subscripts in binary notation. For example, for three qubits we would have
{\small
\begin{equation} \label{eqn:state-vec-3-qubits-binary-notation-transposed}
	|\psi\> = \left[ \begin{matrix}
	    \alpha_{000} &
        \alpha_{001} &
        \alpha_{010} &
        \alpha_{011} &
        \alpha_{100} &
        \alpha_{101} &
        \alpha_{110} &
        \alpha_{111}
	\end{matrix}
    \right]^T
\end{equation}
}
To apply a single qubit gate $U_i$ to the $i$th qubit, one must apply the matrix product (\ref{eqn:mtx-product-update-psi}) to each amplitude whose index differs in the $i$th bit \cite{julich-long-thesis}:
\begin{align} \label{eqn:state-vec-partitioning-stride}
	\alpha_{*\cdots*0_i*\cdots*} = q_{11} \alpha_{*\cdots*0_i*\cdots*} + 
                                   q_{12} \alpha_{*\cdots*1_i*\cdots*} \nonumber \\
    \alpha_{*\cdots*1_i*\cdots*} = q_{21} \alpha_{*\cdots*0_i*\cdots*} + 
                                   q_{22} \alpha_{*\cdots*1_i*\cdots*}
\end{align}
where the subscript $i$ in the index on either 0 or 1 denotes that it is in the $i$th position and the asterisks denote values that are the same on either side of the equation. With this method, only the four entries of $Q_i$ are required to be stored in comparison to the $2^{2n}$ entries of $U$. 

The state vector of $2^n$ amplitudes still needs to be formed, but it can be partitioned across multiple processors. In particular, we use $2^k$ processors for integer $k = 0, 1, 2, ...$ so that each processor must store $2^{n - k + 4}$ bytes in memory. This enables the processing of quantum circuits with significantly more qubits. See Figure \ref{fig:mem-requirements} for the memory requirements. 

\begin{figure}
	\centering
    \hspace*{-2em}
    \includegraphics[scale=0.38]{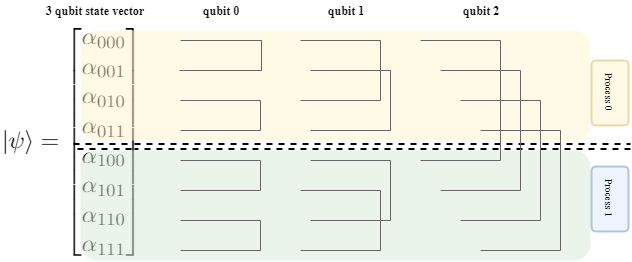}
    \caption{Diagram illustrating the amplitude pairs that need to be updated simultaneously in applying a single qubit gate to a $n = 3$ qubit state vector. If we used two processors, no communication is required for qubits 0 and 1 because the amplitude pairs lie on the same processor. For qubit 2, amplitude pairs on different processors need to be updated simultaneously, so communication is required. The stride between amplitude pairs is $2^i$ for the $i$th qubit.}
    \label{fig:statevector}
\end{figure}

The cost for this increased performance is more involved communication between processors. See Figure \ref{fig:statevector} for an illustration of three qubits with 2 processors. Here, the first four amplitudes are stored on processor 0 and the last four are stored on processor 1. In applying a gate to the qubit zero, no communication between processors is required because the stride between amplitudes as per (\ref{eqn:state-vec-partitioning-stride}) is only one, and these elements lie on the same process. A similar situation holds for qubit one. However, for qubit two, amplitudes on different processors need to be updated simultaneously and so communication is required. 

In general, processor 0 holds the first $2^{n - k}$ amplitudes, processor 1 holds the next $2^{n - k}$ amplitudes, and so on until the last processor which holds the last $2^{n - k}$ amplitudes. Communication between processors is not required for qubits $0, 1, ..., n - k - 1$ and is required for qubits $n - k, n - k + 1, ..., n - 1$. When no communication is required, processors update their amplitudes in parallel for efficient gate application. For qubits that require communication, the processor pairs need to be determined and a communication protocol must be implemented. 


\subsection{\label{sec:single-node-implementation-svp}Single Node Implementation}

\begin{figure}
	\centering
    \includegraphics[scale=0.65]{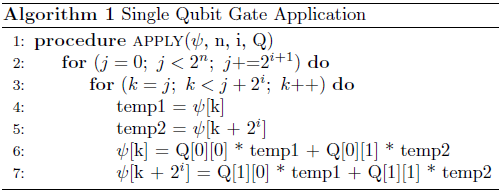}
    \caption{Algorithm to apply a single qubit gate. A multiple (two) qubit gate with control qubit $c$ and target qubit $t$ is identical except the updating scheme only occurs on amplitudes with indices where the $c$th bit is one.}
    \label{fig:apply}
\end{figure}

We first implement the state vector partitioning method on a single node. The algorithm \cite{qhipster} shown in Figure \ref{fig:apply} applies the amplitude updating scheme (\ref{eqn:state-vec-partitioning-stride}).
Here, the outer loop goes over all amplitudes in the state vector that are a distance $2^{i + 1}$ apart where the unitary $Q$ is to be applied to the $i$th qubit. The inner loop starts at a value and determines its pair according to the stride $2^i$; it continues in this fashion until it reaches an amplitude that has already been paired with a previous amplitude. Compare Figure \ref{fig:statevector} for the $n = 3$ case on qubits 0, 1, and 2. We remark that a two qubit gate with control qubit $c$ and target qubit $t$ can be implemented in an almost identical manner. The algorithm and update equation \eqref{eqn:state-vec-partitioning-stride} are the same except we only perform the operation on amplitudes with indices where the $c$th bit is one.

\subsubsection{Performance}

\begin{figure}
	\centering
    \includegraphics[scale=0.50]{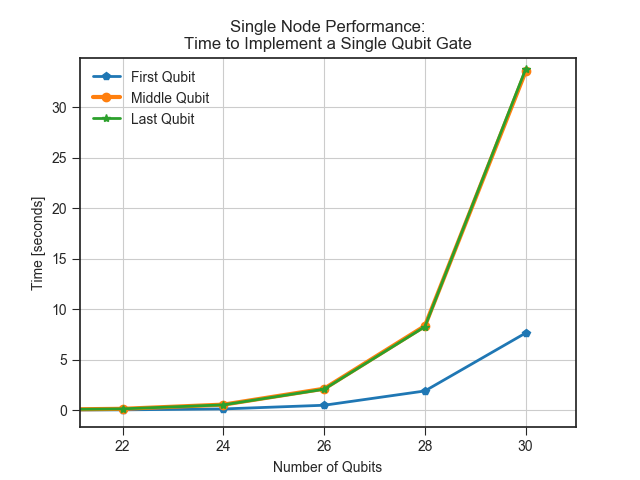}
    \caption{Single node performance results for applying a single qubit gate (no OpenMP parallelization). Different curves show applying gates to different qubits in the circuit. For circuits of size smaller than 24 qubits, single gate application takes less than 0.55 seconds for all curves.}
    \label{fig:svp-singlenode}
\end{figure}

By storing only the state vector of $2^{n + 4}$ bytes, on a single node of 128 GB = $2^{37}$ bytes we should be able to simulate up to a maximum of 33 qubits. Our results for a single gate application (Pauli-$X$) are shown in Figure \ref{fig:svp-singlenode}. We are able to simulate up to $n = 30$ qubits with this method, significantly more than the matrix vector multiplication method using one-dimensional partitioning. For up to $n = 24$ qubits, single qubit gate application takes less than one second. We report results for applying gates on three different qubits in the circuit. As can be seen, the results for each are fairly similar|for small circuits, the curves are identical then begin to diverge slightly as the number of qubits increases. In the maximum size circuit simulated, applying a gate to the zeroth qubit takes 7.64 seconds as compared to 33.55 seconds for the middle qubit and 33.78 seconds for the last qubit. 

\subsubsection{Multi-Threading}

The single node implementation could be optimized further by utilizing multiple threads to carry out the amplitude updating procedure described in Algorithm 1. There are two loops in this algorithm, the ``outer'' loop indexed by $j$ and the ``inner'' loop indexed by $k$. With this nested parallelism we try parallelizing the inner loop and outer loop as well as collapsing the two via the OpenMP \texttt{collapse(2)} directive.

For qubits with low indices in the circuit, the outer loop has many iterations and the inner loop only has few. It is thus expected that parallelizing the outer loop would yield better results in this case. The opposite is true for qubits with larger indices. For qubits in the middle of the circuit, both loops have roughly the same amount of iterations, and one could expect that the collapse directive would perform the best in this case.

Our speedups are shown in Figure \ref{fig:single-node-speedups-table}. As can be seen, these expectations are \textit{generally} true, but the sequential version is found to perform the best in all cases|all speedups are below one. One may try parallelizing both the inner and outer for loops as well as several other methods to overcome data dependencies and attain a speedup. The direct methods tested here proved that the sequential version performed the fastest.

\begin{figure}
	\centering
    { \tiny
    \begin{tabular}{|c|c|c|c|c|c|c||c|} \hline
    		& \multicolumn{3}{c}{\textbf{Inner}} & \multicolumn{3}{c}{\textbf{Outer}}  \\ \hline
            Qubit & $0$ & $n / 2$ & $n - 1$ & $0$ & $n / 2$ & $n - 1$  \\ \hline
            \textbf{1 thread}  & 0.049 & 0.967 & 0.215 & 0.933 & 0.975 & 0.972 \\ 
            \textbf{2 threads} & 0.008 & 0.493 & 0.035 & 0.189 & 0.483 & 0.997  \\ 
            \textbf{4 threads} & 0.007 & 0.481 & 0.032 & 0.491 & 0.771 & 0.987  \\ 
            \textbf{8 threads} & --    & 0.617 &  --   & 0.120 & 0.371 & 0.957  \\ \hline 
    \end{tabular} 
    \vspace*{0em}
    \begin{tabular}{|c|c|c|c||}
    				& \multicolumn{3}{c}{\textbf{Collapse}} \\ \hline 
    	Qubit 		& 0  	& $n / 2$ 	& $n - 1$ 	\\ \hline 
        1 thread 	& 0.996 & 0.977 	& 0.992 \\ 
        2 threads	& 0.595 & 0.877 	& 0.505 \\ 
        4 threads   & 0.292 & 0.471 	& 0.611 \\ 
        8 threads   & 0.149 & 0.654 	& 0.729 \\ \hline 
    \end{tabular}
    }
    \caption{Tables showing the speedups on a $n = 28$ qubit circuit with three different parallelization methods|parallelizing the inner for loop (inner), parallelizing the outer for loop (outer), and using a collapse directive (collapse). The sequential and parallel times are for a single qubit gate application and the speedup is the ratio of these. Speedups are shown for the zeroth qubit, the middle qubit, and the last qubit in the circuit for different numbers of threads.}
    \label{fig:single-node-speedups-table}
\end{figure}

\subsection{\label{sec:multinode}Multinode Implementation}

\begin{figure}
	\centering
    \includegraphics[scale=0.45]{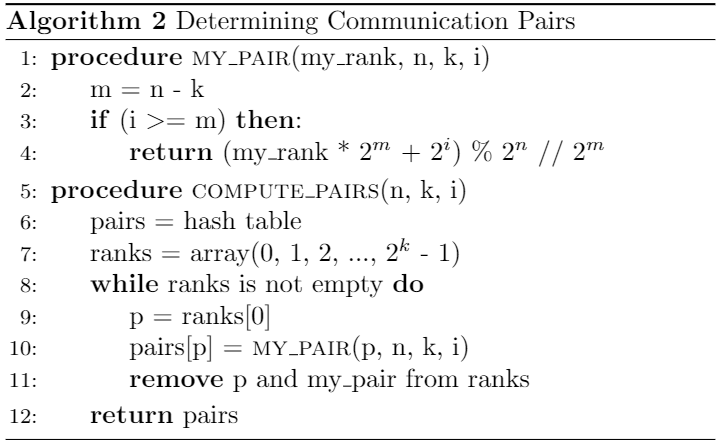}
    \caption{Algorithm for determining pairs of processors that need to communicate with each other.}
    \label{fig:commpairs}
\end{figure}

We now partition the state vector $|\psi\>$ among $2^k$ processors with each storing $2^{n - k}$ local amplitudes. As mentioned, qubits with index $i < n - k$ require no communication. For later qubits, communication pairs must be determined and a communication protocol must be established. Our algorithm for determining communication pairs is shown in Figure \ref{fig:commpairs}.

First, a function is declared to compute the communication pair of a single processor with given rank. This is achieved by computing the global index of the amplitude needed for communication then determining which process it lies on by dividing by the size of local arrays. Note that the double slash \texttt{//} indicates floor division. Next, a function is written to compute the pairs of all given processes by iterating through an array of all ranks in increasing order and computing the corresponding pair. In our implementation, a processors rank and it's pair are stored in a hash table for easy and efficient access. The processor rank and pair index are then removed from the array of ranks and the process continues until all processors have a pair. Note that iterating through the ranks in increasing order is essential for the correctness of this algorithm. If, for example, we started at process 1 instead of process 0 with 3 qubits, 4 processors $(k = 2)$, and $i = 2$, then the communication pair of process 1 would be incorrectly determined to be process 2. This is because the stride $2^i$ always takes the current process to the process with the next rank. By starting at process 0, iterating through the ranks in increasing order, and removing pairs as they are computed, this algorithm produces the correct communication pairs.
\begin{figure*}
	\centering
	\includegraphics[scale=0.4]{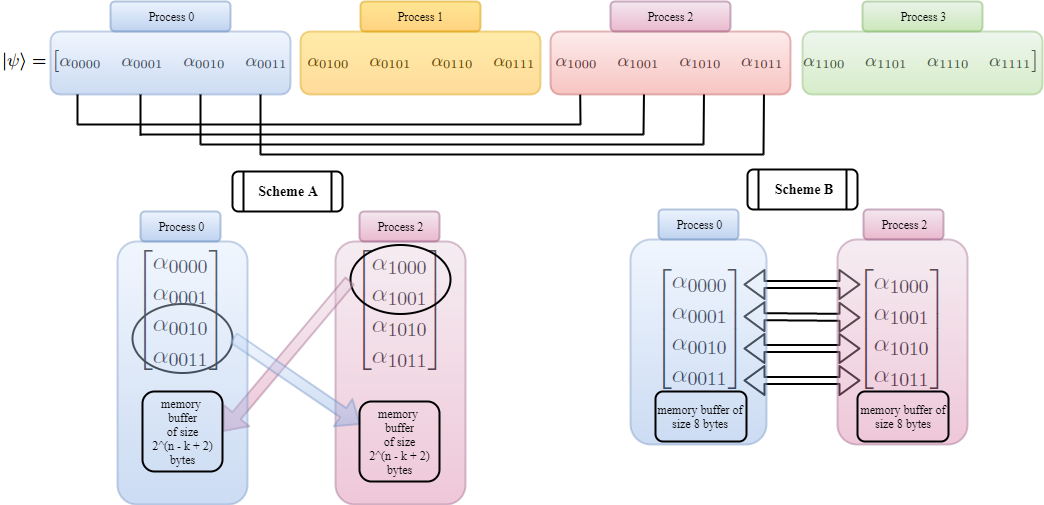}
    \caption{Two different communication protocols for updating amplitudes via the state vector partitioning method when communication is required. Here we show $n = 4$ qubits distributed across $2^k = 4$ process applying a single qubit gate to the third qubit ($i = 3$). Amplitudes that need to be updated simultaneously are determined by the stride $2^i$ and communication pairs are found by Algorithm 2. In Scheme A \cite{julich-long-thesis}, processor 2 sends the first half of its local state vector to process 0, and process 0 sends the latter half of its state vector to process 2. Both processors compute the respective new amplitudes via \ref{eqn:state-vec-partitioning-stride} and send the updated information back. In Scheme B, processors communicate single amplitudes as they iterate through the amplitude updating algorithm (Algorithm 1). This method requires significantly more communication overhead but allows for simulation of larger circuits.}
    \label{fig:comm_protocols}
\end{figure*}

\subsubsection{Communication Protocol}

Once pairs have been established, communicating the correct amplitudes is simple because arrays on all nodes share the same local indices. We implement a different communication protocol than \cite{julich-long-thesis} in which processors send half their local arrays to their processor pairs, compute the amplitudes, then send the arrays with updated amplitudes back. This strategy requires an additional memory buffer of $2^{(n - k + 4) - 2}$ bytes on each node. This space reserved for communication limits the total number of qubits that can be simulated. Our implementation sends only the required amplitudes at each iteration in Algorithm 1. This method allows for larger quantum circuits to be simulated at the cost of increased communication between processors and thus decreased performance. See Figure \ref{fig:comm_protocols}.

Our communication scheme has an additional advantage if relatively few processors are used. Define the \textit{communication ratio} 
\begin{equation} \label{eqn:comm-ratio}
	c \equiv \frac{n - k}{k}
\end{equation}
to be the number of qubits that do not require communication to the number of qubits that do. For example, if $n = 30$ qubits and 8 processors are used so that $k = 3$, then communication is required only for qubits 27, 28, and 29 and $c = 9$. In an average circuit, there will thus be nine times as many gates that can be performed without communication. The first communication protocol limits the number of qubits that can be stored on a single node due to the necessary memory buffer, thus requiring more processors for equally sized circuits, increasing the number of qubits that require communication and decreasing the communication ratio $c$. If $k$ increases from $3$ to 5, for example, the $c$ roughly halves from 9 to 5. With this scheme, only there will only be five times as many gates that can be performed without communication. In short, this scheme handles communication more efficiently but inherently requires more communication. In this respect, our protocol has the twofold benefits of increasing the number of qubits that can be simulated by limiting the necessary memory buffer and increasing circuit simulation performance by maximizing $c$. 

\subsubsection{Performance}

With our multi-node implementation, we are able to simulate up to $n = 33$ qubits distributed across $2^6$ processors with a best single gate application time of 3.04251 seconds. This time comes from a qubit where no communication is required and nodes can update in parallel. Although our communication protocol allows for larger circuits to be simulated with fewer qubits requiring communication, it significantly increases the time to apply a unitary to a qubit where communication is needed. This is because amplitudes are sent pairwise in each iteration of the amplitude updating algorithm. 
\begin{figure}
	\centering
    \hspace*{-2em}
    \includegraphics[scale=0.4]{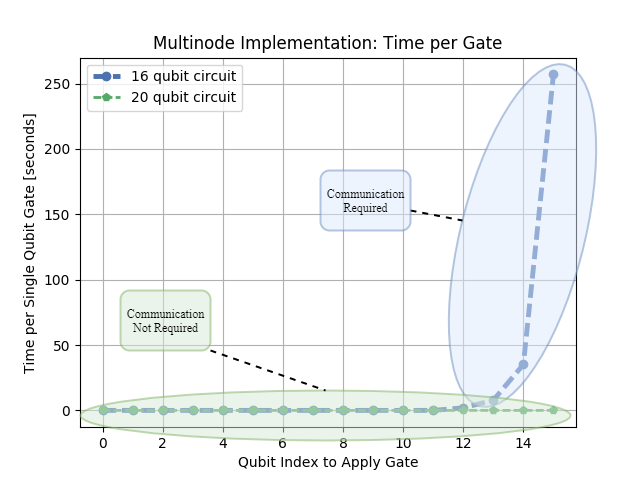}
    \caption{A plot showing the time per single qubit gate application for qubits with different indices in the distributed memory implementation using a single amplitude communication protocol. Here, data is shown for 16 and 20 qubits distributed across 8 processors. The time to apply a gate increases significantly for qubits where communication is required. For clarity, the communication time in the 20 qubit circuit is not shown: the time increases from around 0.002 seconds to 1700 seconds with this communication scheme.}
    \label{fig:multinode}
\end{figure}

Shown in Figure \ref{fig:table-of-simulator-benchmarks} are the performances of some other recently implemented distributed memory quantum circuit simulators as well as simulators using other methods.
\begin{figure}[h!]
	\centering { \scriptsize 
    \begin{tabular}{|c|c|c|c|c|} \hline
        {\tiny\textbf{Simulator}} & {\tiny\textbf{Computer}} & {\tiny\textbf{Benchmark}} & {\tiny\textbf{Max Qubits}} & {\tiny\textbf{Time [s]}} \\ \hline
        This Work & Laconia   & SQG & 33 & 3.04   \\ \hline
        QX		  & unknown       & ENT      & 29 & 63.02   \\ \hline 
        LIQ$U|\>$ & desktop       & ENT      & 23 & 4.09    \\ \hline
        qH$i$PSTER  & Stampede    & SQG & 40 & 1.22    \\ \hline 
        Ref \cite{45qubit-simulation} & Cori II & SQG & 45 & 0.97 \\ \hline
    \end{tabular}
    }
    \caption{Benchmarks for several quantum simulators|SQG stands for single qubit gate and ENT stands for entanglement. The performance of QX and L$i$Q$U|\>$ are recorded from \cite{QX,adv-sim-compression-overview-2017} and the performance of qH$i$PSTER is recorded from \cite{qhipster}.}
    \label{fig:table-of-simulator-benchmarks}
\end{figure}
The simulator qH$i$PSTER was run on the Stampede supercomputer at the Texas Advanced Computing Center at the University of Texas at Austin. At number 10 in the current Top500, it consists of 6400 compute nodes each with two sockets of Xeon ES-2680 connected with QPI, each node having 32 GB DDR4 memory. Using 1024 nodes, 40 qubits were able to be simulated for an average single qubit gate time of 1.22 seconds. The simulator in \cite{45qubit-simulation} was performed on the Cori II supercomputer with 8192 compute nodes and 0.5 petabytes of memory and was able to simulate up to 45 qubits. 


\section{\label{sec:future-work}Future Work and Proposed Improvements}

\subsection{Improvements on Communication Protocols}

The communication scheme implemented in this paper is desirable for maximizing the number of qubits able to be simulated by minimizing extra memory buffer requirements, and for maximizing the communication ratio of the quantum circuit|the ratio of the number qubits that do not require communication to the number that do. With this method the largest possible number of qubits in the circuit need the minimum time to perform a single qubit gate. For 33 qubits partitioned among $2^5$ processors, our maximum number achieved, only qubits with indices $i \ge 28$ need communication; the first 28 need no communication and gates can be performed in seconds. 

The problem with this implementation is the large increase in time needed for gates that do require communication. One may consider, in addition to the communication ratio $c$, the ratio of time per gate application with communication vs without communication,
\begin{equation} \label{eqn:comm-time-ratio}
	T \equiv \frac{t_{\textrm{comm}}}{t_{\textrm{no comm}}}.
\end{equation}
In our performance tests, we found that for a circuit with $n = 16$ qubits, $t_{\textrm{no comm}} \approx 10^{-4}$ seconds whereas $t_{\textrm{comm}} \approx 10^2$ seconds so that $T$ is on the order of $10^6$. Other simulators \cite{qhipster} report results such that $T \approx 10^2$. 

It is desirable to have high $c$ and low $T$ for efficient distributed memory simulations of quantum circuits, but the two are (roughly) inversely proportional. An improvement to both the Scheme A communication and Scheme B communication protocols in Figure \ref{fig:comm_protocols} may be to combine the best aspects of both and meet somewhere in the middle, communicating some number of amplitudes between 1 (Scheme B) and $2^{n - k - 2}$ (Scheme A) to both decrease communication overhead and minimize the extra memory buffer.

\subsection{Non-Sequential Storage of the State Vector}

Consider the scenario in Figure \ref{fig:statevector}. If we stored the amplitudes of the state vector $|\psi\>$ in the following non-increasing order
{ \tiny
\begin{equation*}
	|\psi\> = \left[ \begin{matrix}
	    \alpha_{000} &
        \alpha_{001} &
		\alpha_{100} &
        \alpha_{101} &
        \alpha_{010} &
        \alpha_{011} &
        \alpha_{110} &
        \alpha_{111}
	\end{matrix}
    \right]^T,
\end{equation*}
}
or in decimal notation for clarity
\begin{equation*}
	|\psi\> = \left[ \begin{matrix}
	    \alpha_0 &
        \alpha_1 &
		\alpha_4 &
        \alpha_5 &
        \alpha_2 &
        \alpha_3 &
        \alpha_6 &
        \alpha_7
	\end{matrix}
    \right]^T,
\end{equation*}
then process 0 holds amplitudes with indices 0, 1, 4, and 5 and process 1 holds amplitudes with indices 2, 3, 6, and 7. This means that there is no longer any communication required for qubit two, the last qubit in the circuit, because all the amplitudes that require simultaneous updating are now on the same processor. However, we get no free lunch as qubit one now requires communication between processors. (Qubit zero still needs no communication.)

In effect we have interchanged or relabeled qubit two and qubit three. This may seem useless at first, but by doing so we can assign communication to qubits that have the smallest number of gates applied to them. Equivalently, we can minimize communication for the ``most active'' qubits in the circuit. If many gates are acting on the later qubits in the circuit|the ones that require communication|then the performance of a sequential storage simulator will be particularly poor. However, if the amplitudes of the state vector are stored in a more clever (non-sequential) fashion before being partitioned to different processors, it can be made so that the later qubits no longer require communication (but some other qubits do), and performance can be improved. This would change the stride for the $i$th qubit and may make implementation slightly more complicated, but it is a hardware and software independent optimization that could be utilized.


\section{\label{sec:conclusions}Conclusions}

In this paper we have successfully simulated up to $n = 33$ qubits on the MSU Laconia supercomputer using multi-node and multi-threaded parallelization. Our communication scheme has the benefit of increased communication ratio and ability to simulate larger circuits at the cost of high gate application on qubits requiring communication. We introduced a non-sequential storage of the state vector $|\psi\>$ to map communication needs to the most idle qubits in a circuit. In our multi-node implementation, we presented an efficient algorithm for determining the communication pairs amongst processors. In our single node implementation, we successfully simulated up to 30 qubits with all qubits of index 24 or lower requiring less than 0.55 seconds per gate. Further work could be done to improve upon the communication protocol presented in this paper and decrease the overall time per gate application.

Though we have focused on high performance computing optimizations, recently several quantum simulators \cite{adv-sim-compression-overview-2017,qtorch} have been implemented that rely on data compression rather than distributed computing. These work by compressing information stored and the state vector and representing quantum gates in a more efficient way, similar to the advantage of the state vector partitioning method vs the naive parallel matrix vector multiplication in this paper. These simulators boast of being able to compete with those run on the fastest computers in the world when they are run on a standard personal computer. Both are interesting solutions to the exponential memory problem incurred in trying to simulate a quantum computer. As quantum computers with exponential performance improvements over classical machines slowly become a reality, it is interesting to see how high performance computing methods can compete with the nascent machines to push back (or perhaps halt altogether) the era of ``quantum supremacy.'' This computing competition is useful for both paradigms in making new discoveries and bringing about new computational methods and techniques.


\section{\label{sec:acknowldegements}Acknowledgements}

This work was supported in part by Michigan State University through computational resources provided by the Institute for Cyber-Enabled Research. RL acknowledges support from an Engineering Distinguished Fellowship from Michigan State University. 


\section{\label{sec:references}References}

\end{document}